\documentstyle[12pt,epsfig]{article}

\begin{document}
\date{}
\title{
               Maximum likelihood analysis of the first KamLAND results
}
\date{}
\author{      A.\ Ianni$^{a}$      }
\maketitle
\begin{center} {\em
a) INFN - Laboratori Nazionali del Gran Sasso,
\\   S.S. 17bis Km 18+910, I-67010 Assergi (Aquila), ITALY}
\end{center}
\date{}

\begin{abstract}
A maximum likelihood approach has been used to analize the first results from KamLAND emphasizing the
application of this method for low statistics samples. The goodness of fit has been determined exploiting
a simple Monte Carlo approach in order to test two different null hytpotheses. It turns out that with the present
statistics the neutrino oscillation hypothesis has a significance of about 90\% 
(the best-fit for the oscillation parameters from KamLAND are found to be: $\delta m_{12}^2 \sim 7.1 \times
10^{-5}$~eV$^2$ and $\sin^2 \theta_{12} = 0.424/0.576$), while the no-oscillation hypothesis of
about 50\%. 
Through the likelihood ratio the hypothesis of no disappearence is rejected at about 99.9\% C.L. with the present data from
the positron spectrum. 
A comparison with other analyses is presented.                
\end{abstract}

\section{Introduction}
\label{Sec1}

The purpose of this paper is to perform a likelihood analysis of the first KamLAND
results~\cite{kamland}. KamLAND is a reactor-based neutrino oscillation experiment with a baseline
(source-detector distance) larger than 100~km~\cite{bemporad}. 
This allows KamLAND to explore with
a terrestrial anti-neutrino beam part of the oscillation parameters space which is of interest for
solar neutrinos. In particular, KamLAND can test the Large Mixing Angle (LMA) solution for the solar neutrinos
puzzle~\cite{bahcallbook}. 

Experimental measurements~\cite{homestake,gallex,sage,superk,sno} of the solar 
neutrinos flux on Earth seems to show that these particles undergo matter-enhanced flavor transformations 
(MSW effect)~\cite{msw,bahcallbook}. However, a global analysis which takes into account only the solar neutrinos
measurements cannot identify a unique solution for the parameters which drive the oscillations~\cite{fogli,bahcall}.
Including the new results from KamLAND the scenario changes and only one possibility, namely the mentioned LMA
solution, survives~\cite{fogli,bahcall,nunokawa,barger,balantekin,smirnov,maltoni,valle}. 

In the following we present
a new analysis of the KamLAND data. This analysis, although performed with the first results shows few features such
as the importance of the systematic uncertainties which may affect future data treatments.

\begin{figure}[t!]\begin{center}
\epsfig{bbllx=80pt,bblly=60pt,bburx=900pt,bbury=420pt,height=14truecm,
        figure=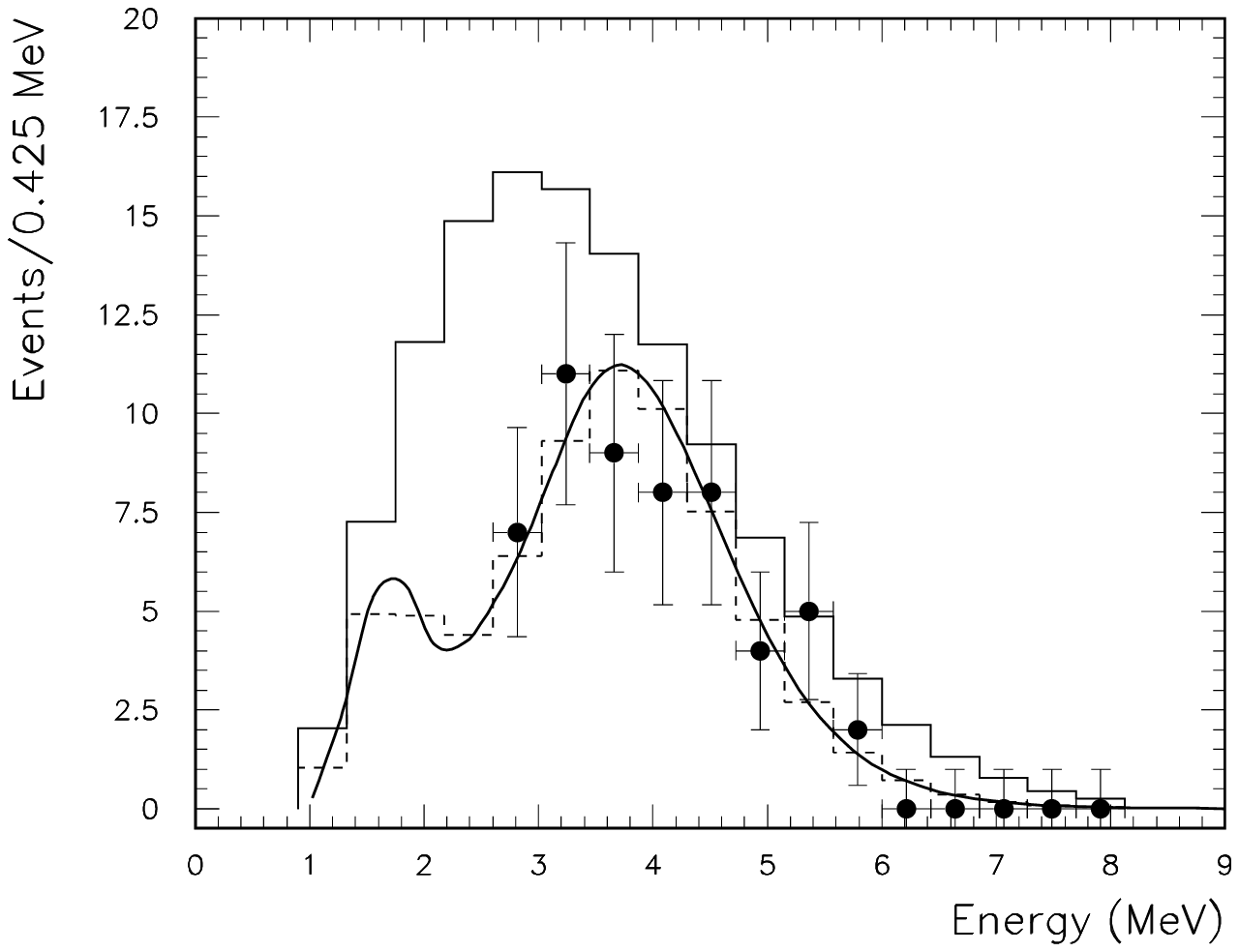}%
\caption{\label{fig1}\footnotesize
	KamLAND results (dots). Solid line (hystogram): expected number of events according to
	an average normalization of 86.8 events and no oscillations. Dashed line: expected distribution with
	oscillations and using the ML fit. Solid line: best-fit ML with oscillations.}
\end{center}\end{figure}

In Fig.~\ref{fig1} we show the kamLAND results as from~\cite{kamland}.
It can be noticed that the statistics is rather poor at the moment and 5 bins have zero entries. Therefore, a least square
 analysis of the data could be not appropriate. In this case, as for low counting experiments, 
the method of analysis commonly used is that of the Maximum Likelihood (ML).
So, in this paper we attempt to perform a ML analysis. 
In order to define our likelihood function we introduce some definitions. 
We call, for the experimental KamLAND results, $N^{obs}_{tot}$ and $N^{obs}_i$ the total number of 
observations and the number of entries in the bin ith, respectively. 
Moreover, we assume that  $N^{obs}_{tot}$ is a Poisson random variable. Hence, the
joint p.d.f. for the set of data shown in Fig.~\ref{fig1} is the product of Poisson distributions~\cite{book}. 
\begin{equation}
L({\bf{v}}) = \prod_{i=1}^N \frac{({N^{th}_i(\bf{v}}))^{N^{obs}_i}}{N^{obs}_i!}Exp({-N^{th}_i(\bf{v}})) \, \label{eq1}
\end{equation} 
where $N^{th}_i$ is the expected number of entries in the bin ith and $\bf{v}$ is the vector of unkown parameters we wish to
estimate through the ML method. According to the ML prescription in order to estimate the unknown parameters we should maximize
the likelihood function. This is usually done through the log-likelihood function which in the 
large-N limit\footnote{Here, we call $N$ the data sample size.} (i.e. with high statistics) is parabolic. 
For the log-likelihood we write
\begin{equation}
ln L({\bf{v}}) = -\int_{a}^{b} f(x;{\bf{v}}) dx + 
\sum_{i=1}^N N^{obs}_i ln( \int_{x{_i}}^{x_{i+1}} f(x;{\bf{v}}) dx) 
\, \label{eq2} 
\end{equation}
with $f(x;{\bf{v}})$ being the function which describes the physical process under investigation
\footnote{Here, $N_i^{th}= \int_{x{_i}}^{x_{i+1}} f(x;{\bf{v}}) dx$.} and, $a$ and $b$ 
define the region of interest for the random variable $x$, which is measured. 

Once the parameters have been estimated using the log-likelihood the difficult task is to determine the confidence intervals for
the same parameters. In the large-N limit this is done through the covariance matrix using the second derivatives of the
log-likelihood function or through the relation~\cite{book,hannam}
\begin{equation}
ln L({\bf{v}}) = ln L_{max} - \frac{Q}{2} 
\, \label{eq3} 
\end{equation}
where $Q$ defines the condifence region as a function of the number of parameters. Values of $Q$ are tabulated in~\cite{book} as
an example. Eq.~(\ref{eq3}) can also be used when the likelihood function is not Gaussian, i.e. in the small-N
limit. In this case, however, the classical definition of confidence region is only approximated by eq.~(\ref{eq3}). Depending on how
accurately the uncertainties should be reported one could try to estimate the level of the approximation 
by a Monte Carlo~\cite{book}. For a multidimensional likelihood an alternative approach for confidence regions estimation is to maximize the
log-likelihood function with respect to all parameters but one. The profile of $ln L_{max}$ against this latter can be used as a
one dimensional problem and, as an example, Q=1 in eq.~(\ref{eq3}) will corresponds to a 68.3\% confidence interval.

Finally, once the parameters and their uncertainties have been estimated, a goodness-of-fit calculation can be carried out through the
classical Pearson's $\chi^2_{Per}$  test, where
\begin{equation}
\chi^2_{Per} = \sum_{i=1}^N \frac{(N^{obs}_i - N^{th}_i)^2}{N^{th}_i}. \label{eq4} 
\end{equation}
Eq.~(\ref{eq4}) follows a $\chi^2$ distribution only in the large-N limit and the rule of thumb is that the number of entries in
the experimental histogram should be such that $N^{obs}_i > 5$~\cite{book,rpp}. When the large-N limit is not reached a Monte Carlo study 
to determine the statistics of $\chi^2_{Per}$ should be performed. Only in this way a correct P-value can be
calculated.

\section{Analysis of the first KamLAND results}
\label{Sec2}

Recently, KamLAND results have been analized~\cite{nunokawa,fogli,bahcall,barger,balantekin,smirnov,maltoni,valle} and used to estimate the
solar neutrino oscillation parameters performing a global analysis including solar neutrinos and
short-baseline experiments. In particular, in~\cite{nunokawa,fogli,bahcall,barger,balantekin} KamLAND results have
been analized cosidering the statistics: $\chi^2_P = -2log \lambda_P$\footnote{The implementation of this $\chi^2$ is suggested by
the Review of Particle Properties~\cite{rpp}.}, where $\lambda_P$ is a normalized likelihood function given
by the product of Poisson distributions~\cite{book}.  So, $\lambda_P$ and the likelihood used in this paper only differ by a factor and
should have the same best-fit parameters. Yet, the confidence region estimation requires a Monte Carlo
calculation when the
statistics is poor because only in the large-N limit $\chi^2_P$ follows a $\chi^2$ distribution with $N-p$
d.o.f., being $p$ the
number of estimated parameters~\cite{book,rpp}. In~\cite{smirnov,maltoni,valle} a different approach for the statistical analysis has been
used. Here, a multidimensional $\chi^2$-function is used with the covariance matrix calculated from the experimental
errors shown in Fig.~\ref{fig1} and the systematic uncertainty from~\cite{kamland}. Of course, the use of a
$\chi^2$-function implies the assumption of Gaussian errors. As pointed out in Sec.~\ref{Sec1} this could be not appropriate
for the data set of KamLAND.

In the following we have used the above considerations and the ML method to analize the KamLAND results and compare the  findings
with the other methods already implemented.

For KamLAND and in a 2$\nu$ scenario we write
$$
N^{th}_i(\alpha,\delta m^2, \sin \theta_{12}) = 
\alpha A \int_{E_i}^{E_{i+1}} dE \int dE' R(E,E') \sigma(E_{\bar{\nu}_e}) 
\phi(E_{\bar{\nu}_e}) \Big(\sum_i \frac{P_i}{d^2_i} P_i^{ee}\Big) 
$$
\begin{equation}
 = \int_{E_i}^{E_{i+1}} dE N^{th}(E,\alpha,\delta m^2, \sin \theta_{12}) \, \label{eq5}
\end{equation}
where $E$ and $E'$ are the measured and real visible energy, respectively,
and $E_{\bar{\nu}_e} = E' + 0.8$~MeV the energy of the incoming $\bar{\nu}_e$. Moreover, A is a normalization factor which 
accounts for the number of target protons, the detection efficiency, the data taking time
and conversion of units, $\sigma(E')$ is the cross-section for the inverse 
$\beta$-decay~\cite{sigma}, $P_i$ is the thermal power (in units of GW) of the ith reactor and $d_i$
its distance from the KamLAND detector~\cite{bemporad}, $\phi(E')$
\footnote{This flux takes into account the differential spectrum
of the $\bar{\nu}_e$'s, the average energy and the intensity fraction of each fuel component.} is the 
$\bar{\nu}_e$'s flux at the detector, weighed over the different fuel
components~\cite{bemporad,rotunno} in U and Pu as from~\cite{kamland} and calculated using the differential
spectrum from~\cite{vogel}. 
In eq.~(\ref{eq5}), $R(E,E')$ is a Gaussian energy resolution function with $\sigma = 0.075
\sqrt{E'(MeV)}$~\cite{kamland}. 

In order to determine the constant A, we have normalized eq.~(\ref{eq5}) by using the 
information and the average expected number of
events (86.8) from~\cite{kamland}. Yet, we introduce a correction factor, $\alpha$, which accounts 
for the systematic uncertainties.
The $\bar{\nu}_e$'s survival probability, $P^{ee}_i$, is written
\begin{equation}
P^{ee}_i = 1 - \sin^2 2\theta_{12} \sin^2 \Big( 1.27 \frac{\delta m^2(eV^2) d_i (m)}{E_{\bar{\nu}_e} (MeV)} \Big) \, \label{eq6}
\end{equation}
In particular, to take into account the sistematic uncertainties we have multiplied the likelihood function in eq.~(\ref{eq1}) 
by a Gaussian given as a function of $\alpha$ with mean value equal to one and $\sigma_{sys}=0.064$,
the total systematic uncertainty quoted in~\cite{kamland}. So, the log-likelihood function to maximize looks like
$$
ln L(\alpha, \delta m^2, \sin \theta_{12}) = - \int_{2.6 MeV}^{8.125 MeV} dE N^{th}(E,\alpha,\delta m^2, \sin
\theta_{12}) +
$$
\begin{equation} 
+ \sum_{i=1}^{13} N^{obs}_i ln \int_{E_i}^{E_{i+1}} dE N^{th}(E,\alpha,\delta m^2, \sin \theta_{12}) - 
 \frac{1}{2}\Big(\frac{\alpha-1}{\sigma_{sys}}\Big)^2 \, \label{eq7} 
\end{equation}
with $N^{th}$ from eq.~(\ref{eq5}).

\begin{figure}[t!]\begin{center}
\epsfig{bbllx=80pt,bblly=60pt,bburx=900pt,bbury=420pt,height=14truecm,
        figure=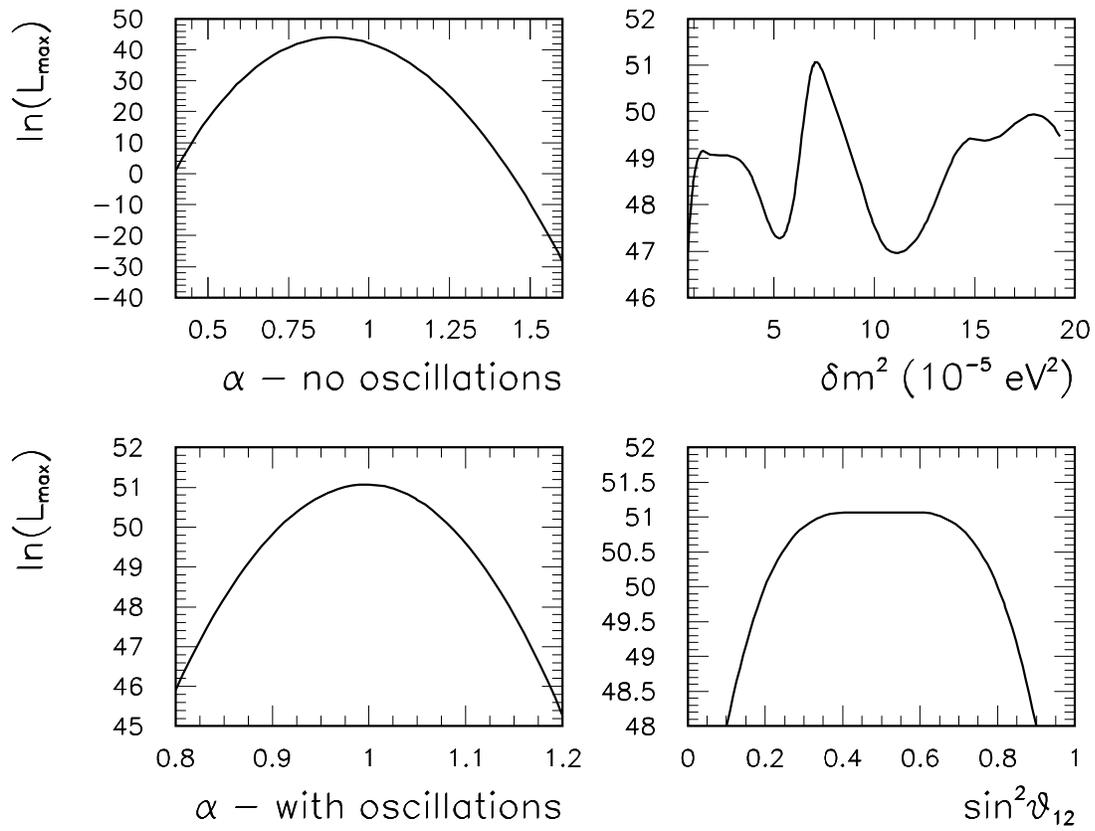}%
\caption{\label{fig2}\footnotesize
	Profiles of $ln L_{max}$ against the parameters. Up-right: no-oscillation hypothesis.}
\end{center}\end{figure}

We have searched for maxima of $ln L$ from eq.~(\ref{eq7}) using two hypotheses. 
The first assumes no-oscillation and the only non-zero
parameter is $\alpha$. In this case the best-fit is for $\alpha=0.892$. We point out that without the systematic
uncertainty the ML method gives $\alpha \sim 0.6$, which corresponds to the ratio between the measured and
expected number of events~\cite{kamland}. The number of events in the absence of
oscillations is $77.4^{+4.4}_{-2.5}$ with 1$\sigma$ confidence interval according to eq.~(\ref{eq3}). 
This integrated rate is in agreement within 1$\sigma$ with the number
used for the absolute normalization: 86.8$\pm$5.6 expected events as from~\cite{kamland}. The second hypothesis assumes oscillations
according to eq.~(\ref{eq6}). In this case we have found several local maxima and two global ones symmetric with respect to
$\sin^2 \theta_{12} =0.5$. The best-fit points are: ($\alpha$, $\sin^2 \theta_{12}$, $\delta m^2$) =
(0.997, $^{0.576}_{0.424}$,7.11), 
where $\delta m^2$ is given in units of 10$^{-5}$~eV$^2$. To determine the confidence regions
for the parameters we
have studied the profile of $ln L$ against one parameter while maximizing with respect to the others. 
The result of this study around the global maxima is shown in Fig.~\ref{fig2}. 
It can be noticed, in particular looking at the profiles of $\alpha$, that the log-likelihood is not
parabolic. So, though small there is a deviation from the large-N behavior of the data under
investigation. The 68.3\% confidence intervals, from eq.~(\ref{eq3}), are reported in Tab.~\ref{tab1}.
   
\begin{table}[hbt]\setlength{\tabcolsep}{1.5pc}
\caption{\em Confidence intervals at 68.3\%. Best-fit values are (0.997,$^{0.576}_{0.424}$,
7.11$\times$10$^{-5}$~eV$^2$).  \label{tab1}}
\begin{center}\begin{tabular}{|l|c|c|c|}
\hline
Model  & $\alpha$ & $\sin^2 \theta_{12}$ & $\delta m^2$ \\
       &          &                      & (10$^{-5}$~eV$^2$) 
                                                                      \\ \hline
No-oscil.   & [0.836,0.949] &  -   &  - \\ \hline
oscil. & [0.936,1.057] & [0.253,0.747]   & [6.72,7.67] \\ \hline
\end{tabular}\end{center}\end{table}

\begin{figure}[t!]\begin{center}
\epsfig{bbllx=80pt,bblly=60pt,bburx=900pt,bbury=420pt,height=14truecm,
        figure=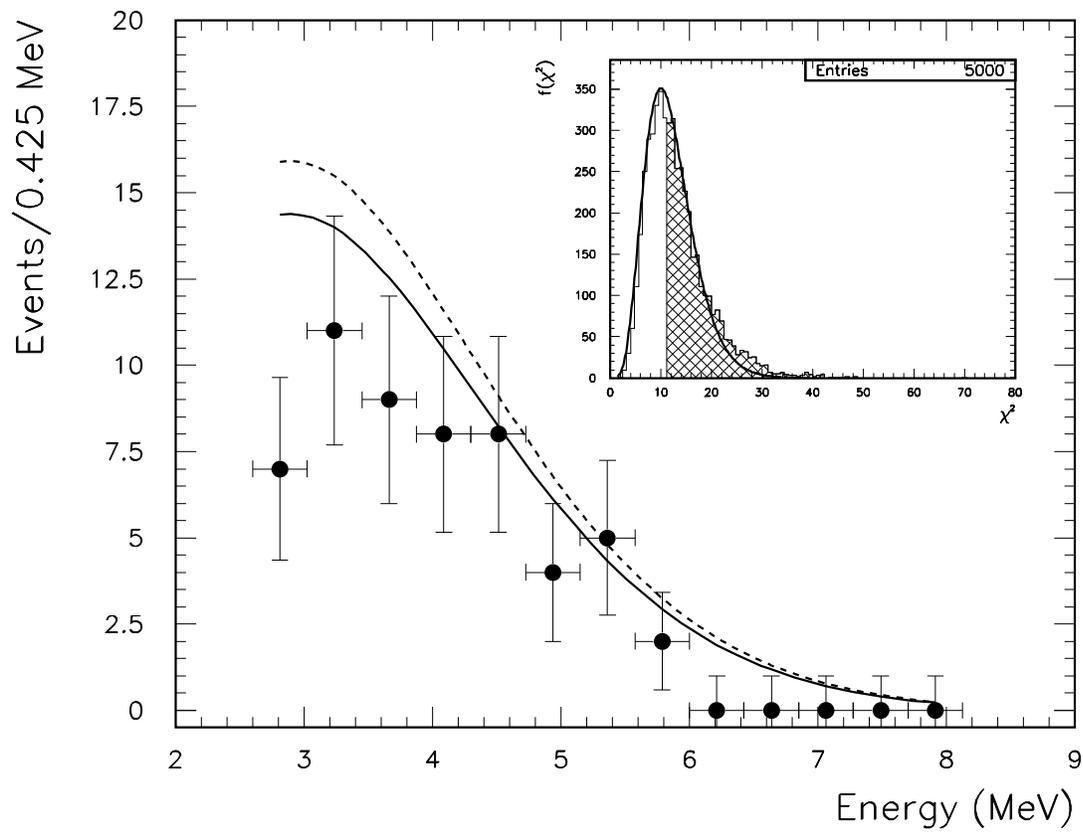}%
\caption{\label{fig3}\footnotesize
	Best fit with the no-oscillation hypothesis. Up-left: $\chi^2_{Pers}$ statistics. 
	The dashed line corresponds to a total systematic uncertainty reduced at 2\%. See text for details.}
\end{center}\end{figure}

\begin{figure}[t!]\begin{center}
\epsfig{bbllx=80pt,bblly=60pt,bburx=900pt,bbury=420pt,height=14truecm,
        figure=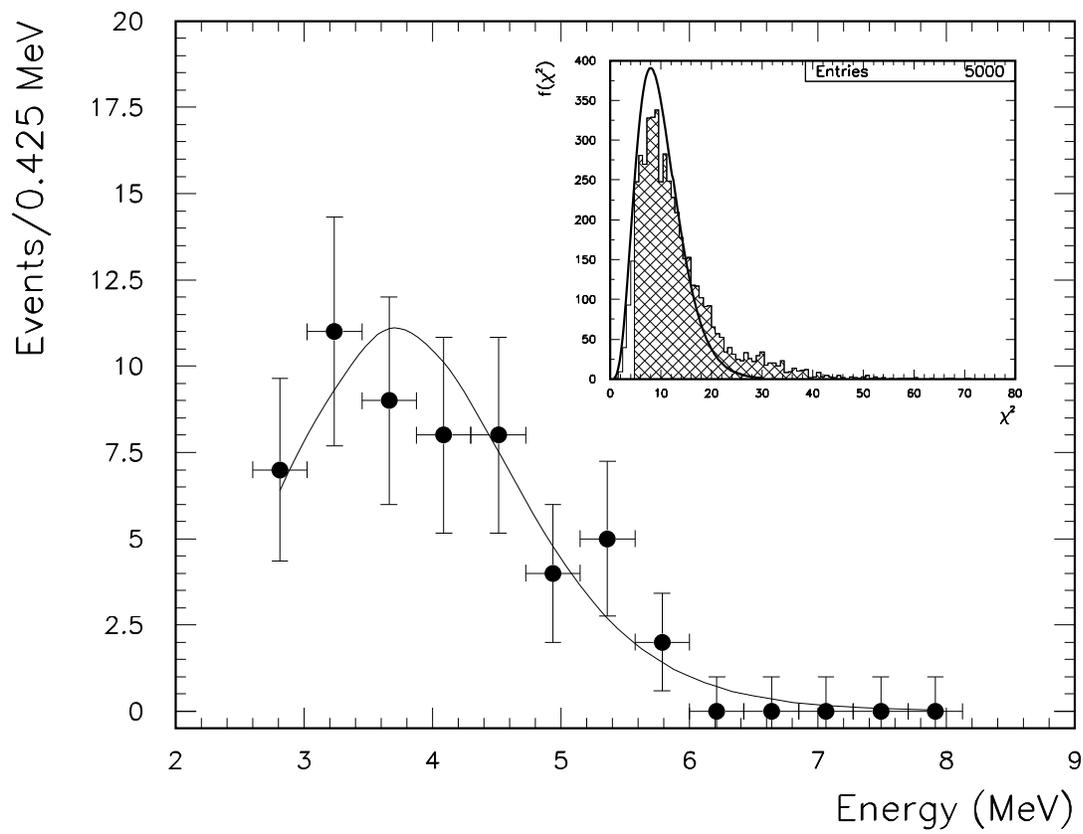}%
\caption{\label{fig4}\footnotesize
	Best fit with the oscillation hypothesis. Up-left: $\chi^2_{Pers}$ statistics. See text for details.}
\end{center}\end{figure}

\begin{figure}[t!]\begin{center}
\epsfig{bbllx=80pt,bblly=40pt,bburx=500pt,bbury=500pt,height=12truecm,
        figure=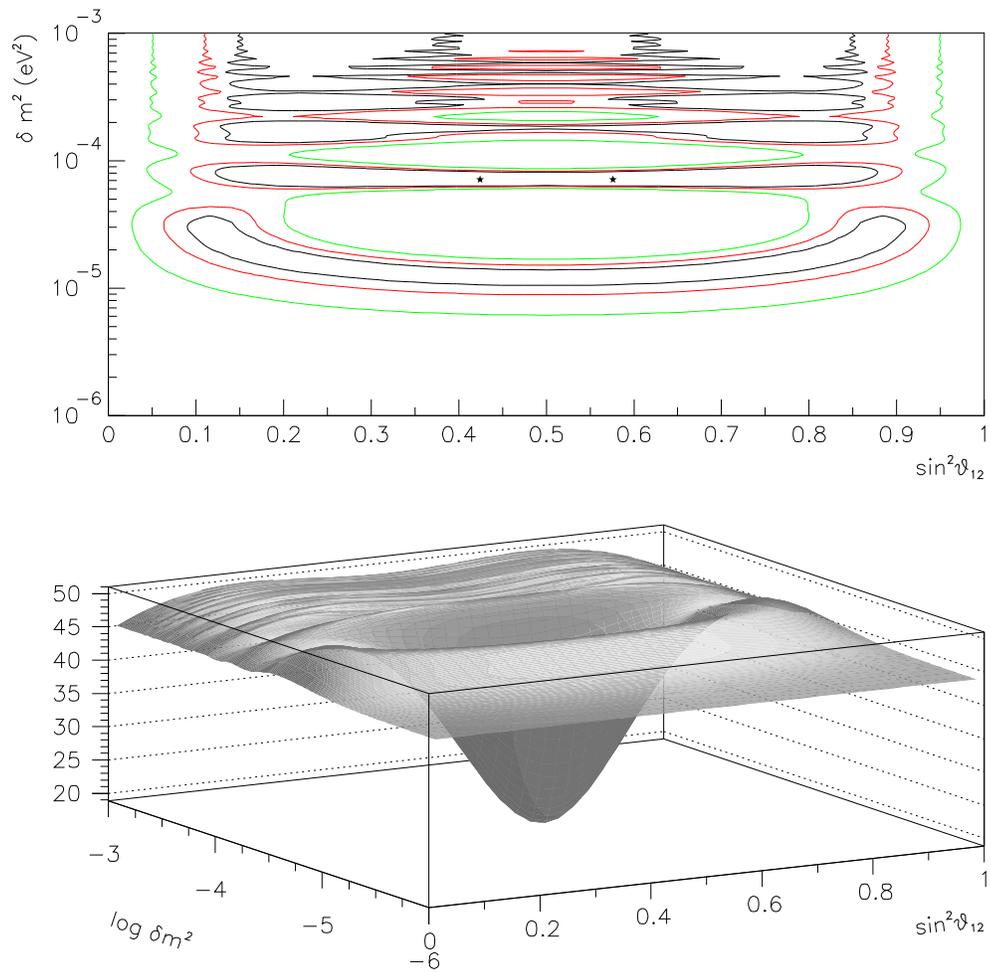}%
\caption{\label{fig5}\footnotesize
          Maximum likelihood confidence regions at 90\%, 95\% and 99\% (upper plot) for two-flavor active neutrino 
	  oscillations at KamLAND. The best-fit points are indicated by black dots (upper plot). Profile of
	  ln L (lower plot) found by maximizing with respect to $\alpha$ (see text for details).}
\end{center}\end{figure}

In order to study the goodness-of-fit for the two hypotheses under consideration we have calculated the distribution of
$\chi^2_{Per}$ by generating Poisson values for N$^{obs}_i$ based on the mean value for N$^{th}_i$ according to the fit performed
with the ML method. For the no-oscillation hypothesis we show in Fig.~\ref{fig3} the KamLAND data against the ML fit. On the
up-right corner we also show the distribution of $\chi^2_{Per}$ together with that of a $\chi^2$ p.d.f. with 12 d.o.f. 
The darkened area show the fraction of the distribution above the measured value for $\chi^2_{Per}$.
 In this
case the P-value is 53\% ($\sim$48\% using the $\chi^2$ distribution).
Following the same reasoning in Fig.~\ref{fig4} we show the ML fit assuming oscillations. Again in the up-right corner we
report the $\chi^2_{Per}$ statistics. The g.o.f is 93\% ($\sim$90\% using the $\chi^2$
distribution). For completeness in Fig.~\ref{fig1} we show the best-fit curve assuming oscillations
together with the distribution of the expected events in the standard scenario (with $\alpha=0$).
Finally, in order to compare our analysis with the others on the KamLAND results, in Fig.~\ref{fig5} we 
show the 90\%, 95\% and 99\%
confidence regions for the oscillation parameters (from eq.~(\ref{eq2})) together with the the profile of 
$ln L(log(\delta m^2),\sin^2 \theta_{12})$. This latter has been determined by maximizing with respect to $\alpha$.

\section{Conclusions}
\label{Sec3}

In conclusion we have analized the first results of KamLAND through the ML method for a 
no-oscillation hypothesis and an oscillation one. The former gives a P-value of about 93\% and 
the latter of about 53\%. By chance we get a result in perfect agreement with what reported in~\cite{kamland} 
about how the positron spectrum is consistent with a neutrino oscillations assumption although we have a
slightly different oscillation parameters. 
Furthermore, we also are in perfect agreement with the conclusion in~\cite{kamland} about a no-oscillation 
assumption for the spectrum analysis. However, in the paper this finding is clearly due to the present
statistics and the systematic uncertainty. On the contrary, the method of analysis in~\cite{kamland} is not well explained.
 We should also remind that the strongest evidence in favour of neutrino oscillations 
comes from the ratio between the measured and expected number of events~\cite{kamland}. 
In this paper we have used the information from the positron spectrum and in order to properly
make a comparison between the two hypotheses discussed (oscillation and no-oscillation) we have worked out the ratio 
$\lambda = L_{max}^{no-osc}/L_{max}^{osc}$ (likelihood ratio) which turns out to be $8.45\times 10^{-4}$.
This small value gives an indication that the observed positron spectrum in KamLAND 
rules out the no disappearance scenario with the present statistics. Computing -2$ln\lambda$, which follows a $\chi^2$
distribution with 2 d.o.f. in this case, it turns out that the assumption which restrict the number of parameters, i.e.
the no disappearence hypothesis, is rejected at the level of about 99.9\%.

For the sake of completeness we have reduced the systematic uncertainty at 2\% to study the trend of the fit. As shown
in Fig.~\ref{fig3} the fit gets worse for the no-oscillation hypothesis and the new P-value is equal to about 30\%. 

We have also shown that a $\chi^2$ test, in this we could call quasi low-count rate 
experimental scenario, gives within few \%'s a result in agreement with that reported using a Monte Carlo 
calculation for $\chi^2_{Per}$. 

A limiting point of the analysis presented and common with others on the same argument is the
treatment of the systematic uncertainty which are combined in two main sources in~\cite{fogli} and in one correction factor to the overall
normalization in~\cite{nunokawa,bahcall,barger,balantekin} and here. Moreover, no matter effects 
are taken into account here. Yet, this correction gives only a
small contribution~\cite{bahcall,rotunno}. 

\section*{Aknowledgments}

A.I. is greatful to C. Giunti, E. Lisi and D. Montanino for discussions during
the preparation of this paper.

\newpage

\newpage

\newpage

\newpage

\newpage

\end{document}